\documentclass[journal,transmag]{IEEEtran}

\usepackage{cite}
\usepackage{graphicx}
\graphicspath{{figures/}}
\DeclareGraphicsExtensions{.pdf}
\usepackage{amsmath}
\usepackage{booktabs}
\usepackage{siunitx}
\usepackage{url}
\usepackage{xcolor}

\usepackage[nohyperlinks]{acronym}
\usepackage[hidelinks]{hyperref}
\acrodef{DTAP}{dual-term air-gap permeance}
\acrodef{FEM}{finite element method}
\acrodef{UMP}{unbalanced magnetic pull}
\acrodef{IM}{induction motor}
\acrodef{MEC}{magnetic equivalent circuit}
\acrodef{MMF}{magnetomotive force}
\acrodef{NISM}{numerical integration solution method}
\acrodef{AISM}{analytical integration solution method}
\acrodef{RMS}{root-mean-square}
\acrodef{HAM}{hybrid analytical model}

\begin{document}

\title{Improved Closed-Form Identification of the Air-Gap Permeance in Magnetic Equivalent Circuits of Rotor-Eccentric Electric Motors}

\author{\IEEEauthorblockN{Philip~Desenfans,~\IEEEmembership{Member,~IEEE}\IEEEauthorrefmark{1,2}}
\IEEEauthorblockA{\IEEEauthorrefmark{1}Department of Electrical Engineering (ESAT), KU Leuven, Bruges, Belgium},
\IEEEauthorblockA{\IEEEauthorrefmark{2}Flanders Make @ KU Leuven, Leuven, Belgium}%
\thanks{Corresponding author: P. Desenfans (e-mail: philip.desenfans@kuleuven.be).}}

\IEEEtitleabstractindextext{%
\begin{abstract}
Fast and accurate simulation of electric motors is essential for their design, fault diagnosis and control. \Acfp{MEC} offer a computationally efficient alternative to the \acf{FEM}, yet their flux tube representation poorly captures the air-gap field. This creates a challenge, as the air gap stores most of the magnetic energy and governs the electromagnetic torque and, in rotor-eccentric motors, the \acf{UMP}. Existing air-gap permeance expressions vary widely in their results for torque ripple and \ac{UMP} and account for tooth-normal eccentricity only. This paper proposes the \acf{DTAP} model, a closed-form expression for the air-gap permeance between a stator and a rotor tooth that accounts for tooth-normal and tooth-tangential eccentricity, with and without rotor skew. Its parameters are identified from a \ac{FEM} model of the air gap and the complete set of tooth heads, which includes the shielding by the neighbouring teeth. For three motors, \ac{DTAP} reproduces the \ac{FEM} air-gap permeance with a 2.8--4.8 times smaller error than the state of the art, at a comparable computational cost. Implemented in the \ac{MEC} of an eccentric induction motor, it reproduces the torque ripple within 1.6\,\% and the peak \ac{UMP} within 1.3\,\% of the reference, i.e., with 2.8 and 27 times smaller deviations than the state of the art. \ac{DTAP} thereby enables \acp{MEC} to predict the torque pulsations and bearing loads of eccentric motors more accurately without sacrificing computational speed.
\end{abstract}

}

\maketitle
\IEEEdisplaynontitleabstractindextext
\acresetall

\section{Introduction}
\IEEEPARstart{T}{he} \ac{MEC} method was introduced in the 1960s for the modelling of electrical machines~\cite{laithwaite1967, carpenter1968}. It represents the magnetic topology by a lumped network of reluctances, permeances and \acp{MMF}, which discretises the machine far more coarsely than the \ac{FEM} and is therefore solved much faster. \acp{MEC} have since been developed for induction~\cite{ostovic1986, ostovic1988}, salient-pole synchronous~\cite{slemon1990}, permanent-magnet~\cite{rasmussen1997}, switched reluctance~\cite{kokernak2000} and axial-flux machines~\cite{hemeida2019}, extended to 3-D~\cite{amrhein2009}, and used for design~\cite{amrhein2010}, loss computation~\cite{tariq2010}, fault analysis~\cite{naderi2017, forstner2020, lannoo2020}, hardware-in-the-loop emulation~\cite{tavana2016, fleming2016} and control~\cite{forstner2020}.

Owing to the low permeability of air relative to the core, the magnetic behaviour of an electrical machine is largely governed by its air gap~\cite{ge2022}. In a \ac{MEC}, the air gap is represented by permeances between stator and rotor, which vary with the rotor angle and displacement. As the torque and \ac{UMP} follow exclusively from the partial derivatives of these permeances~\cite{ostovic1989, lannoo2020}, the air-gap model largely determines the accuracy of the \ac{MEC}.

Air-gap models for \acp{MEC} can be grouped into five classes: (i) overlap, (ii) network-regeneration, (iii) sliding-interface, (iv) \ac{HAM}, and (v) \ac{FEM}-based methods. The \textit{overlap} method, introduced by Ostovi\'{c}~\cite{ostovic1989}, derives the permeance from the overlap between facing teeth, either smoothed by retaining the first two Fourier terms~\cite{mahyob2007, fleming2016, li2021angular} or used directly~\cite{sudhoff2007, jayarajan2022magnetic}. Mesh-based \acp{MEC} discretise the air gap more finely into elementary network elements and couple the rotating and stationary meshes either by regenerating overlap permeances at every rotor position~\cite{amrhein2009, han2016unstable, tong2018quasi} or by interpolating the magnetic scalar potential at a sliding interface~\cite{asfirane2019scalar, hemeida2024magnetic}. \acp{HAM} replace the air-gap network by an analytical field solution, such as the Fourier subdomain method~\cite{hemeida2019} or conformal mapping~\cite{rezaee2021improved, rezaee2023modified}. However, the overlap method neglects fringing between teeth that do not overlap, mesh-based methods require a fine air-gap mesh, and \acp{HAM} simplify the air-gap geometry, e.g., by neglecting slotting.

\ac{FEM}-based methods, to which this work belongs, capture slotting and fringing by fitting an expression to \ac{FEM} evaluations of the air-gap permeance, e.g., a trigonometric polynomial~\cite{delforge1995}, a piecewise constant and quadratic function~\cite{gyselinck2013}, or a geometry-based double-Gaussian expression~\cite{lannoo2020}, which was later extended to rotor skew~\cite{desenfans2024} and to a segment of constant permeance~\cite{silva2023}. Three limitations remain. First, the \ac{FEM} evaluation differs among authors, e.g., in the surface over which the flux is integrated, and often omits the neighbouring teeth, so that reported accuracies are difficult to compare. Second, while the existing characterisation of tooth-normal eccentricity fails above 50\,\% of the nominal air-gap length~\cite{desenfans2024}, tooth-tangential eccentricity has so far been neglected. Yet both components always occur together: an eccentric rotor is displaced normal to the stator teeth aligned with the direction of eccentricity and tangential to those perpendicular to it. Third, a comprehensive comparison of the existing methods is lacking, especially regarding the predicted torque ripple and \ac{UMP}.

This work proposes the \ac{DTAP} model, which provides closed-form expressions for the air-gap permeance and its partial derivatives, fitted to the \ac{FEM}.\footnote{The \ac{FEM} identification, the fitting and the evaluation of \ac{DTAP} and of the five published methods it is compared with are implemented in the Python package APIT (Air-gap Permeance Identification Tool), available at \url{https://gitlab.kuleuven.be/m-group-campus-brugge/wavecore_public/phd-philip-desenfans/apit4mec}.} Compared with the state of the art, (i) the \ac{FEM} evaluation considers all teeth and the entire air gap, thereby including the shielding effect of the neighbouring teeth; (ii) the agreement of the permeance and its partial derivatives with the \ac{FEM} is improved; (iii) tooth-tangential eccentricity is modelled; (iv) rotor skew is included in closed form, together with both eccentricity components; and (v) \ac{DTAP} is compared with five published methods in terms of accuracy, computation time, and the torque ripple and \ac{UMP} predicted by a \ac{MEC}. Sections~\ref{sec:unskewed} and~\ref{sec:permeance_skew} derive \ac{DTAP} for unskewed and skewed rotors, Section~\ref{sec:fem} describes its \ac{FEM} identification, Section~\ref{sec:results} presents the results, and Sections~\ref{sec:limitations} and~\ref{sec:conclusion} discuss the limitations and conclude the paper.

\section{Unskewed Air-Gap Permeance}\label{sec:unskewed}
\begin{figure}[!t]
  \centering
  \includegraphics[width=\columnwidth]{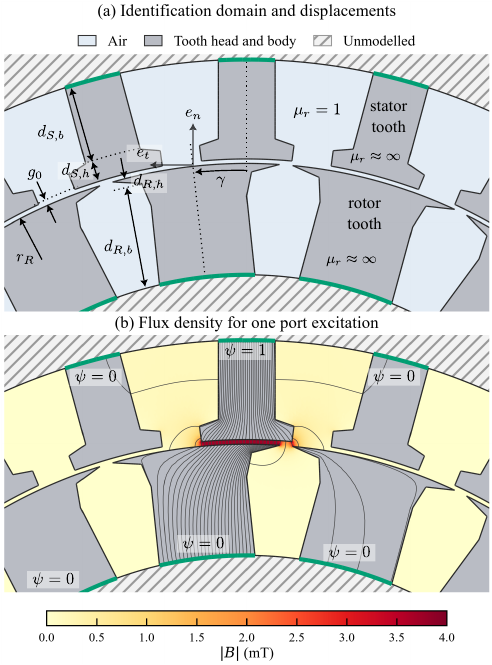}
  \caption{Illustration of notation and air-gap permeance identification method. The used motor geometry is provided in \cite{amrhein2010}.}
  \label{fig:domain}
\end{figure}

Fig.~\ref{fig:domain}(a) introduces the notation. A rotor tooth $j$ is positioned relative to a stator tooth $k$ by their angular separation $\gamma$ and by the displacement components $e_n$ and $e_t$, along and perpendicular to the centreline of stator tooth $k$, respectively. Further, $g_0$ is the nominal air-gap length and $r_R$ the rotor outer radius. The air-gap permeance $\mathcal{P}_{a}$ relates the flux $\Phi$ connecting the teeth to the \ac{MMF} $\mathcal{F}$ across the air gap; its \ac{FEM} evaluation is detailed in Section~\ref{sec:fem}.

Ostovi\'{c}~\cite{ostovic1989} divided the air-gap permeance curve into a plateau during maximal tooth overlap, a raised cosine during partial overlap, and zero thereafter. \ac{DTAP} achieves a similar shape with the inverted cubic Hermite smoothstep
\begin{equation}
S(x;A,B)=1-3f^2+2f^3,\quad f=\text{clip}\left(\frac{|x|-B}{A-B},0,1\right),
\label{eq:smoothstep}
\end{equation}
whose plateau ends at $|x|=B$ and which reaches zero at $|x|=A$. Besides matching the \ac{FEM} more closely than a raised cosine, the smoothstep is a polynomial, so it can be differentiated and integrated in closed form.

\begin{figure}[!t]
  \centering
  \includegraphics[width=\columnwidth]{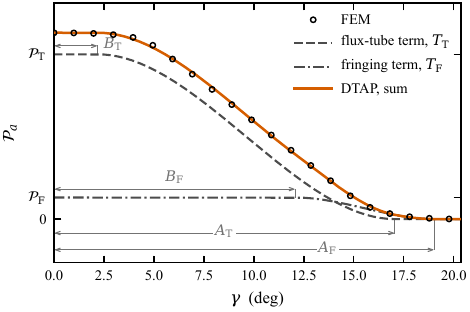}
  \caption{\ac{DTAP} as the sum of two smoothstep functions, with parameters indicated.}
  \label{fig:anatomy}
\end{figure}

\begin{figure}[!t]
  \centering
  \includegraphics[width=\columnwidth]{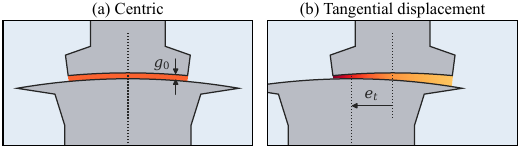}
  \caption{Effect of tangential eccentricity on the distribution of the air-gap length across a tooth pair at $\gamma=e_n=0$.}
  \label{fig:tangential}
\end{figure}

As shown in Fig.~\ref{fig:anatomy}, \ac{DTAP} sums two smoothsteps, hence the name ``dual-term''. For a centric rotor,
\begin{equation}
\mathcal{P}_a = \mathcal{P}_\mathrm{T}S(\gamma; A_\mathrm{T}, B_\mathrm{T}) + \mathcal{P}_\mathrm{F}S(\gamma; A_\mathrm{F}, B_\mathrm{F}),
\label{eq:dtap_unskewed_unecc}
\end{equation}
where $\mathrm{T}$ and $\mathrm{F}$ denote the flux-tube and fringing terms. Their amplitudes $\mathcal{P}$ and parameters $A$ and $B$ are six of the eight fitted parameters of \ac{DTAP}. Under eccentricity, (\ref{eq:dtap_unskewed_unecc}) becomes
\begin{equation}
\mathcal{P}_a = \underbrace{\mathcal{A}_\mathrm{T}S(u;A_\mathrm{T},B_\mathrm{T})\,Q(u)}_{T_\mathrm{T}} + \underbrace{\mathcal{A}_\mathrm{F}S(u;A_\mathrm{F},B_\mathrm{F})}_{T_\mathrm{F}},
\label{eq:dtap_unskewed_ecc}
\end{equation}
with $\mathcal{A}_\mathrm{T}=\mathcal{P}_\mathrm{T}h_\mathrm{T}$ and $\mathcal{A}_\mathrm{F}=\mathcal{P}_\mathrm{F}h_\mathrm{F}$. Normal eccentricity enters through the gap closure ratios
\begin{equation}
h_\mathrm{T}=\frac{g_0}{g_0-e_n}\quad\text{and}\quad h_\mathrm{F}=\frac{g_0}{g_0-\kappa e_n}.
\label{eq:gap_closure_ratio}
\end{equation}
The flux-tube term scales linearly with $h_\mathrm{T}$, as expected from geometry~\cite{ostovic1989}, whereas flux bulging, and hence fringing, decreases with normal eccentricity~\cite{desenfans2024}, which the fitted parameter $\kappa$ captures. As shown in Fig.~\ref{fig:tangential}, a tangential displacement makes the clearance vary across the tooth face. It shifts the curve through the effective separation angle $u=\gamma+e_t/r_R$ and tilts it through
\begin{equation}
Q(u)=1+\varepsilon \frac{u}{A_\mathrm{T}} \quad \text{with} \quad \varepsilon=s\frac{e_t}{g_0-e_n},
\label{eq:tilt_factor}
\end{equation}
where the eighth fitted parameter, $s$, sets the degree of tilt. Setting $e_n=e_t=0$ recovers (\ref{eq:dtap_unskewed_unecc}).

\subsection{Partial Derivatives}

Through virtual work, closed-form partial derivatives of the air-gap permeance allow a fast and numerically stable computation of the torque and \ac{UMP}~\cite{lannoo2020}. With $\mathcal{F}_{kj}$ the \ac{MMF} across the air-gap permeance $\mathcal{P}_{a,kj}$ between stator tooth $k$ and rotor tooth $j$, and $N_S$ and $N_R$ the numbers of stator and rotor teeth,
\begin{align}
T_\mathrm{em} &= \frac{1}{2}\sum_{k=1}^{N_S}\sum_{j=1}^{N_R}\mathcal{F}_{kj}^2\,\frac{\partial \mathcal{P}_{a,kj}}{\partial \gamma}, \label{eq:torque}\\
\mathbf{F}_\mathrm{UMP} &= \frac{1}{2}\sum_{k=1}^{N_S}\sum_{j=1}^{N_R}\mathcal{F}_{kj}^2\left(\frac{\partial \mathcal{P}_{a,kj}}{\partial e_n}\,\mathbf{n}_k+\frac{\partial \mathcal{P}_{a,kj}}{\partial e_t}\,\mathbf{t}_k\right), \label{eq:ump}
\end{align}
where $\mathbf{n}_k$ and $\mathbf{t}_k$ are the unit vectors in the directions of $e_n$ and $e_t$. With $S_\mathrm{T}=S(u;A_\mathrm{T},B_\mathrm{T})$, and likewise $S_\mathrm{F}$, differentiating (\ref{eq:dtap_unskewed_ecc}) gives
\begin{align}
\frac{\partial \mathcal{P}_a}{\partial \gamma} &= \mathcal{A}_\mathrm{T} \left(\frac{\partial S_\mathrm{T}}{\partial u}Q +S_\mathrm{T} \frac{\varepsilon}{A_\mathrm{T}} \right) + \mathcal{A}_\mathrm{F} \frac{\partial S_\mathrm{F}}{\partial u}, \label{eq:unskewed_dgamma}\\
\frac{\partial \mathcal{P}_a}{\partial e_n} &= \frac{h_\mathrm{T}}{g_0}\,T_\mathrm{T}\left(2-\frac{1}{Q}\right)+\frac{\kappa h_\mathrm{F}}{g_0}\,T_\mathrm{F}, \label{eq:unskewed_den}\\
\frac{\partial \mathcal{P}_a}{\partial e_t} &= \frac{1}{r_R} \frac{\partial \mathcal{P}_a}{\partial \gamma} + \mathcal{A}_\mathrm{T} S_\mathrm{T} \frac{u}{A_\mathrm{T}} \frac{s h_\mathrm{T}}{g_0}, \label{eq:unskewed_det}
\end{align}
with $\partial S/\partial u=\mathrm{sgn}(u)(6f^2-6f)/(A-B)$, $\partial h_\mathrm{T}/\partial e_n=h_\mathrm{T}^2/g_0$, $\partial h_\mathrm{F}/\partial e_n=\kappa h_\mathrm{F}^2/g_0$ and $\partial\varepsilon/\partial e_n=\varepsilon h_\mathrm{T}/g_0$. The first term of (\ref{eq:unskewed_det}) reuses (\ref{eq:unskewed_dgamma}), as $e_t$ acts through $u$ as an additional separation angle $e_t/r_R$.

\section{Skewed Air-Gap Permeance}\label{sec:permeance_skew}

Rotor skew reduces the harmonics associated with slotting and is adopted in nearly all squirrel-cage \acp{IM} \cite{desenfans2024}. As axial air-gap flux is negligible in radial-flux motors, the air-gap permeance of a rotor skewed by the angle $\alpha$ over the stack length $L$ follows by integrating along the axial position $z\in[-L/2,L/2]$, with $\gamma$ defined at mid-stack:
\begin{equation}
\mathcal{P}_{a,s}=\int_{-L/2}^{L/2}\mathcal{P}_a\left(\gamma+\frac{z\alpha}{L},e_n, e_t\right)\mathrm{d}z.
\label{eq:P_as_orig}
\end{equation}
As $\mathcal{A}_\mathrm{T}$, $\mathcal{A}_\mathrm{F}$ and $\varepsilon$ do not depend on $z$, substituting $v=u+z\alpha/L$, with the half-sweep $\Delta=\alpha/2$, transforms (\ref{eq:P_as_orig}) to
\begin{equation}
\begin{split}
\mathcal{P}_{a,s} = \frac{L}{2\Delta}\Bigg[&\mathcal{A}_\mathrm{T}\int_{u-\Delta}^{u+\Delta} S_\mathrm{T}(v)\,\mathrm{d}v\\
&+ \frac{\mathcal{A}_\mathrm{T}\varepsilon}{A_\mathrm{T}}\int_{u-\Delta}^{u+\Delta} v\,S_\mathrm{T}(v)\,\mathrm{d}v\\
&+ \mathcal{A}_\mathrm{F}\int_{u-\Delta}^{u+\Delta} S_\mathrm{F}(v)\,\mathrm{d}v\Bigg].
\end{split}
\label{eq:skew_integral_substituted}
\end{equation}
Each integral is a difference of antiderivatives. With the zeroth and first moments $M_0(x)=\int_0^x S\,\mathrm{d}v$ and $M_1(x)=\int_0^x vS\,\mathrm{d}v$, and $\delta M_{i,\mathrm{X}}=M_{i,\mathrm{X}}(u+\Delta)-M_{i,\mathrm{X}}(u-\Delta)$ for $\mathrm{X}\in\{\mathrm{T},\mathrm{F}\}$, this gives
\begin{equation}
\mathcal{P}_{a,s} = \frac{L}{2\Delta}\Big[\mathcal{A}_\mathrm{T}\Big(\delta M_{0,\mathrm{T}}+\frac{\varepsilon}{A_\mathrm{T}}\delta M_{1,\mathrm{T}}\Big)+\mathcal{A}_\mathrm{F}\,\delta M_{0,\mathrm{F}}\Big].
\label{eq:skew_moments}
\end{equation}
The moments follow in closed form, as derived in Appendix~\ref{app:moments}:
\begin{equation}
\begin{split}
M_0(x)&=\mathrm{sgn}(x)\big[p+w\,G_0(f)\big],\\
M_1(x)&=\tfrac{1}{2}p^2+w\big[B\,G_0(f)+w\,G_1(f)\big],
\end{split}
\label{eq:moments_closed}
\end{equation}
with $f$ from (\ref{eq:smoothstep}), $w=A-B$, $p=\min(|x|,B)$, $G_0(f)=f-f^3+\tfrac{1}{2}f^4$ and $G_1(f)=\tfrac{1}{2}f^2-\tfrac{3}{4}f^4+\tfrac{2}{5}f^5$.

\subsection{Partial Derivatives}

As $\mathrm{d}M_0/\mathrm{d}x=S$ and $\mathrm{d}M_1/\mathrm{d}x=xS$, the $\mathrm{T}$-terms of (\ref{eq:skew_moments}) recombine at each stack end into $T_\mathrm{T}$, whereas only the weights of the moment differences change with $e_n$ and $e_t$. This gives
\begin{align}
\frac{\partial \mathcal{P}_{a,s}}{\partial \gamma}&=\frac{L}{2\Delta}\big[\mathcal{P}_a(u+\Delta)-\mathcal{P}_a(u-\Delta)\big], \label{eq:skew_dgamma}\\
\frac{\partial \mathcal{P}_{a,s}}{\partial e_n}&=\frac{L}{2g_0\Delta}\Big[h_\mathrm{T}\mathcal{A}_\mathrm{T}\Big(\delta M_{0,\mathrm{T}}+\frac{2\varepsilon}{A_\mathrm{T}}\delta M_{1,\mathrm{T}}\Big)\nonumber\\
&\qquad\qquad+\kappa h_\mathrm{F}\mathcal{A}_\mathrm{F}\,\delta M_{0,\mathrm{F}}\Big], \label{eq:skew_den}\\
\frac{\partial \mathcal{P}_{a,s}}{\partial e_t}&=\frac{1}{r_R}\frac{\partial \mathcal{P}_{a,s}}{\partial \gamma}+\frac{s h_\mathrm{T}L}{2g_0\Delta}\,\frac{\mathcal{A}_\mathrm{T}}{A_\mathrm{T}}\,\delta M_{1,\mathrm{T}}, \label{eq:skew_det}
\end{align}
where $\mathcal{P}_a(u\pm\Delta)$ denotes (\ref{eq:dtap_unskewed_ecc}) evaluated at $u\pm\Delta$. All three reuse the smoothsteps and moments already evaluated for (\ref{eq:skew_moments}).

\section{Finite Element Method Evaluation}\label{sec:fem}

The eight parameters of \ac{DTAP} are fitted to a 2-D magnetostatic \ac{FEM} model of the complete set of tooth heads, so that every permeance includes the influence of the neighbouring teeth. As shown in Fig.~\ref{fig:domain}, the model contains the air gap, the slot openings and all tooth heads, each carried on its tooth body down to a flux port at depth $d_{S,b}$ for the stator and $d_{R,b}$ for the rotor. Heads and bodies have a linear, quasi-infinite relative permeability $\mu_r=10^6$, so that $\mathcal{P}_a$ represents the air gap alone and the core is left to the \ac{MEC}. The field is described by the magnetic scalar potential $\psi$, with $\mathbf{B}=-\mu\nabla\psi$, which is imposed on each port by a Dirichlet condition, with homogeneous Neumann conditions in between (Fig.~\ref{fig:domain}(b)). The flux $\Phi_q$ that tooth $q$ receives from the air gap and carries off through its port is
\begin{equation}
\Phi_q=\sum_{p\neq q}\mathcal{P}_{pq}\left(\psi_p-\psi_q\right),
\label{eq:ring_network}
\end{equation}
where $\mathcal{P}_{pq}=\mathcal{P}_{qp}$ is the permeance between teeth $p$ and $q$. Exciting port $p$ with $\psi_p=1$ while all other ports remain at $\psi=0$ cancels all potential differences in (\ref{eq:ring_network}) except those with respect to tooth $p$, so that $\Phi_q=\mathcal{P}_{pq}$ for every $q\neq p$. For a stator and a rotor tooth, this yields the air-gap permeance, and for two teeth of the same member, the leakage permeance. All excitations of a rotor pose share one sparse factorisation and are solved as one system with multiple right-hand sides.

$\mathcal{P}_a$ is evaluated for 61 separation angles and 81 eccentricities $(e_n,e_t)$ of an $11\times11$ grid with $\sqrt{e_n^2+e_t^2}\le0.9\,g_0$, i.e., 4941 values per machine. As one solution yields all tooth pairs, symmetry reduces this to 648 \ac{FEM} solutions for Motors~1 and~2 and 2466 for Motor~3. The parameters are fitted by bounded nonlinear least squares with the trust-region reflective algorithm~\cite{branch1999} of SciPy~\cite{virtanen2020}. Table~\ref{tab:motors} lists the geometry and fitted parameters of the three squirrel-cage \acp{IM}, where $\theta_S$ and $\theta_R$ are the angles spanned by the tooth heads, $d_{S,h}$ and $d_{R,h}$ their radial depths, and $d_{S,b}$ and $d_{R,b}$ the modelled depths of the tooth bodies. Motor~1 is machine~1 of~\cite{lannoo2020}, Motor~2 the 3.7-kW machine of~\cite{sudhoff2007}, for which a skew of one stator slot pitch is assumed, and Motor~3 the 500-W machine of~\cite{amrhein2010}.

\begin{table}[!t]
\caption{Relevant Motor Geometry and Fitted \acs{DTAP} Parameters}
\label{tab:motors}
\centering
\begin{tabular}{llccc}
\toprule
 & Unit & Motor 1 & Motor 2 & Motor 3\\
\midrule
\multicolumn{5}{l}{\emph{Relevant Motor Geometry}}\\
$N_S$ & -- & 36 & 36 & 24\\
$N_R$ & -- & 28 & 28 & 17\\
$r_R$ & mm & 30.00 & 51.50 & 37.46\\
$g_0$ & mm & 0.25 & 0.50 & 0.34\\
$\theta_S$ & ° & 8.50 & 7.14 & 10.45\\
$\theta_R$ & ° & 11.15 & 11.79 & 19.65\\
$d_{S,h}$ & mm & 2.00 & 0.20 & 1.59\\
$d_{R,h}$ & mm & 2.00 & 0.50 & 2.48\\
$d_{S,b}$ & mm & 6.00 & 6.00 & 6.00\\
$d_{R,b}$ & mm & 1.00 & 6.00 & 6.00\\
$L$ & mm & 400 & 120 & 63\\
$\alpha$ & ° & 0 & 10.00 & 0\\
\midrule
\multicolumn{5}{l}{\emph{Fitted \acs{DTAP} parameters}}\\
$\mathcal{P}_\mathrm{T}$ & $\mu$H/m & 20.44 & 16.73 & 26.00\\
$A_\mathrm{T}$ & ° & 10.53 & 10.93 & 17.05\\
$B_\mathrm{T}$ & ° & 0.00 & 0.63 & 2.17\\
$s$ & -- & 0.0802 & 0.0496 & 0.0586\\
$\mathcal{P}_\mathrm{F}$ & $\mu$H/m & 4.47 & 2.45 & 3.38\\
$A_\mathrm{F}$ & ° & 12.03 & 12.42 & 19.04\\
$B_\mathrm{F}$ & ° & 7.73 & 7.50 & 12.05\\
$\kappa$ & -- & 0.823 & 0.541 & 0.465\\
\bottomrule
\end{tabular}
\end{table}

\section{Results and Discussion}\label{sec:results}

\ac{DTAP} is compared with the method of Ostovi\'{c}~\cite{ostovic1989}, the \ac{NISM} and \ac{AISM} of~\cite{desenfans2024}, the method of Gyselinck and Sabariego~\cite{gyselinck2013} and that of Silva \textit{et al.}~\cite{silva2023}. The first three are implemented as published, and the last two are extended to normal eccentricity by scaling with $h_\mathrm{T}$, as done in \cite{ostovic1989,lannoo2020,desenfans2024}. The free constants of~\cite{desenfans2024} and~\cite{gyselinck2013}, two and three, respectively, are fitted to the same \ac{FEM} data as \ac{DTAP}.

\subsection{Unskewed Permeance}

\begin{figure*}[!t]
  \centering
  \includegraphics[width=\textwidth]{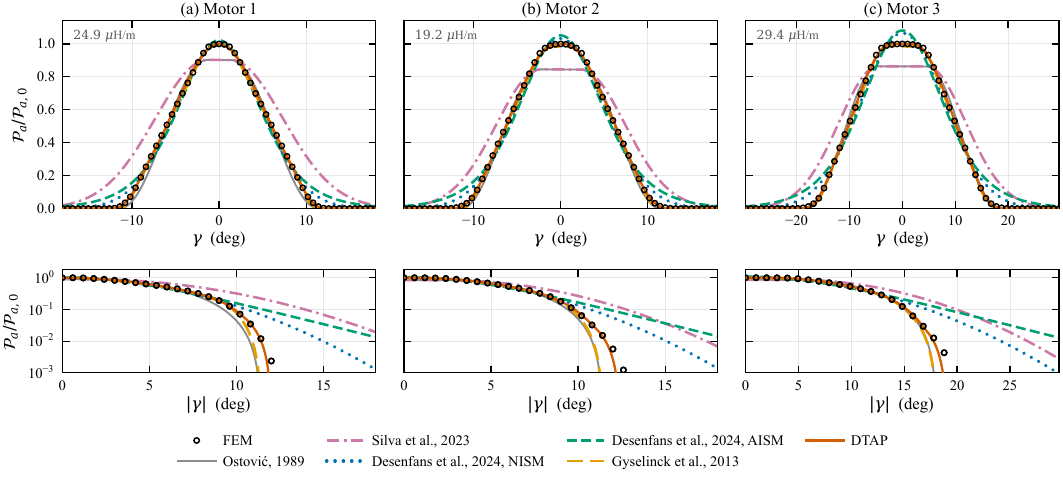}
  \caption{Air-gap permeance at zero eccentricity versus the separation angle $\gamma$ for (a) Motor~1, (b) Motor~2 and (c) Motor~3, normalised by the aligned \acs{FEM} value $\mathcal{P}_{a,0}$.}
  \label{fig:centred}
\end{figure*}

\begin{figure}[!t]
  \centering
  \includegraphics[width=\columnwidth]{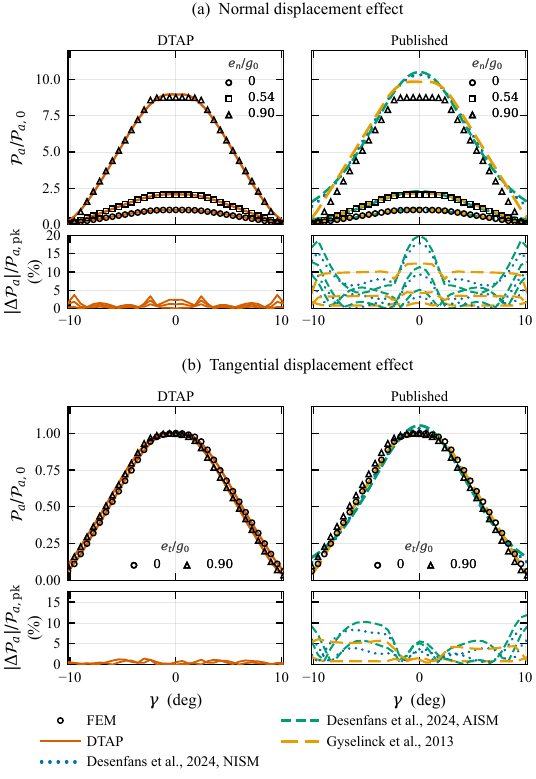}
  \caption{Effect of (a) normal and (b) tangential eccentricity on the air-gap permeance of Motor~2, normalised by $\mathcal{P}_{a,0}$.}
  \label{fig:eccentricity}
\end{figure}

Fig.~\ref{fig:centred} compares all methods without eccentricity. \ac{DTAP} follows the \ac{FEM} most closely, with a mean deviation of 0.47--0.61\,\% of the aligned permeance wherever $\mathcal{P}_a$ exceeds 10\,\% of it, followed by the method of Gyselinck and Sabariego with 0.65--0.96\,\%. The methods of Ostovi\'{c} and Silva \textit{et al.} underestimate the aligned permeance by 9.7--15.5\,\%, and the latter, which neglects the neighbouring teeth that narrow the curve, also overestimates it outside the plateau, giving mean deviations of 4.39--5.71\,\% and 11.54--12.69\,\%, respectively. Lacking a plateau, the \ac{NISM} and \ac{AISM} overestimate the aligned permeance of Motors~2 and~3 by 3.3--8.1\,\%, with mean deviations of 1.48--3.93\,\% and 2.37--5.27\,\%, respectively. In the far tail, the \ac{FEM} permeance vanishes beyond $|\gamma|=12.4$--$19.7^\circ$, because the neighbouring tooth heads intercept the fringing flux. \ac{DTAP} vanishes 0.6--1.0$^\circ$ earlier and the methods of Ostovi\'{c} and Gyselinck and Sabariego 1.1--1.9$^\circ$ earlier, whereas the other methods decay too slowly. Although these deviations appear large on the logarithmic scale, they matter little, as the permeance and its derivatives are small there.

Fig.~\ref{fig:eccentricity} shows the effect of eccentricity on Motor~2 for the three published methods with the lowest mean deviation in Fig.~\ref{fig:centred}. At $e_n=0.9\,g_0$, the aligned \ac{FEM} permeance rises by a factor of 8.76, whereas these methods scale the entire curve by $h_\mathrm{T}=10$ and thus deviate by 9.77--10.06\,\%. As \ac{DTAP} scales its fringing term by $h_\mathrm{F}$ instead, it deviates by only 1.36\,\%. None of the published methods depends on $e_t$, which yields 3.87--4.99\,\% at $e_t=0.9\,g_0$, against 0.59\,\% for \ac{DTAP}.

\begin{table}[!t]
  \centering
  \caption{Relative error of air-gap permeance models against FEM.}
  \label{tab:comparison}
  \setlength{\tabcolsep}{4.5pt}
  \footnotesize
  \begin{tabular}{@{}lcrrr@{}}
    \toprule
    Model & $N$ & Motor 1 & Motor 2 & Motor 3 \\
    \midrule
    Silva et~al., 2023 & \textbf{0} & 14.52\,\% & 12.79\,\% & 13.69\,\% \\
    Ostovi\'{c}, 1989 & \textbf{0} & 4.36\,\% & 5.60\,\% & 5.66\,\% \\
    Desenfans et~al., 2024, AISM & 2 & 3.82\,\% & 5.28\,\% & 6.47\,\% \\
    Desenfans et~al., 2024, NISM & 2 & 2.92\,\% & 4.04\,\% & 4.93\,\% \\
    Gyselinck et~al., 2013 & 3 & 2.18\,\% & 2.99\,\% & 2.59\,\% \\
    DTAP (this work) & 8 & \textbf{0.78\,\%} & \textbf{0.62\,\%} & \textbf{0.72\,\%} \\
    \bottomrule
  \end{tabular}

  \smallskip
  \parbox{\linewidth}{\raggedright $N$: number of fitted constants of the
    model. Lowest value per column in bold.}
\end{table}

Table~\ref{tab:comparison} summarises the mean relative error
\begin{equation}
\eta=\frac{1}{J}\sum_{j=1}^{J}\frac{\big|\mathcal{P}_{a,j}-\mathcal{P}_{a,j}^{\mathrm{FEM}}\big|}{\mathcal{P}_{\mathrm{pk},j}}
\label{eq:mean_error}
\end{equation}
over all $J$ rotor tooth poses of the three machines, where $\mathcal{P}_{\mathrm{pk},j}$ is the \ac{FEM} peak permeance of the eccentricity configuration of pose $j$, which prevents the small permeances in the tail from dominating $\eta$. Points below 1\,\% of $\mathcal{P}_{\mathrm{pk},j}$ are excluded. \ac{DTAP} achieves 0.62--0.78\,\%, which is 2.8--4.8 times lower than the most accurate published method, that of Gyselinck and Sabariego, with 2.18--2.99\,\%. This margin stems mainly from the eccentric poses: at zero eccentricity, the method of Gyselinck and Sabariego deviates by 0.78--1.09\,\% against 0.42--0.53\,\% for \ac{DTAP}, whereas over the eccentric poses it deviates by 2.20--3.01\,\% against 0.63--0.78\,\%. Although \ac{DTAP} has eight fitted constants against at most three, all methods are fitted to the same \ac{FEM} data, so the remaining deviations reflect the limits of their functional forms rather than of their calibration.

\begin{table}[!t]
  \centering
  \caption{Computation time of the unskewed air-gap permeance and its partial derivatives.}
  \label{tab:timing-unskewed}
  \setlength{\tabcolsep}{2.5pt}
  \footnotesize
  \begin{tabular}{@{}lrrrrr@{}}
    \toprule
    Model & $\mathcal{P}_a$ & $\partial\mathcal{P}_a/\partial\gamma$ & $\partial\mathcal{P}_a/\partial e_n$ & $\partial\mathcal{P}_a/\partial e_t$ & Total \\
    \midrule
    Silva et~al., 2023           & 6.28 & 9.88 & 2.87 & N/A & 19.0 \\
    Ostovi\'{c}, 1989            & 5.46 & 10.8 & 0.67 & N/A & 16.9 \\
    Desenfans et~al., 2024, NISM & 4.78 & 9.11 & 2.51 & N/A & 16.4 \\
    Desenfans et~al., 2024, AISM & 7.11 & 8.25 & \textbf{0.11} & N/A & 15.5 \\
    DTAP (this work)             & 6.74 & \textbf{0.92} & 0.84 & 2.11 & 10.6 \\
    Gyselinck et~al., 2013       & \textbf{2.04} & 3.45 & 2.53 & N/A & \textbf{8.02} \\
    \bottomrule
  \end{tabular}

  \smallskip
  \parbox{\linewidth}{\raggedright Derivative times assume $\mathcal{P}_a$ is given. Unit in $\mu$s. N/A: no dependence on $e_t$. Lowest value per column in bold.}
\end{table}

Table~\ref{tab:timing-unskewed} lists the computation time of $\mathcal{P}_a$ and of its partial derivatives for all 1008 tooth pairs of Motor~2, using single-threaded Numba kernels~\cite{lam2015} on an Intel Core i7-1165G7 processor at 2.8\,GHz, as the minimum over 15 repetitions. Derivatives without a published closed form are computed by central differences. The permeance of \ac{DTAP} takes 6.74\,$\mu$s, within the 2.04--7.11\,$\mu$s of the published methods, and its closed-form derivatives take only 0.84--2.11\,$\mu$s, as they reuse its intermediate quantities. Its derivative to $\gamma$ is the fastest of all methods, against 3.45--10.8\,$\mu$s for the central differences and 8.25\,$\mu$s for the closed form of the \ac{AISM}, and its derivative to $e_n$ is faster than the central differences, but slower than the single-factor scalings of Ostovi\'{c} and the \ac{AISM}. In total, including the derivative to $e_t$ that the published methods lack, \ac{DTAP} requires 10.6\,$\mu$s, second only to the method of Gyselinck and Sabariego with 8.02\,$\mu$s.

\subsection{Skewed Permeance}\label{sec:permeance_skew_results}

\begin{figure}[!t]
  \centering
  \includegraphics[width=\columnwidth]{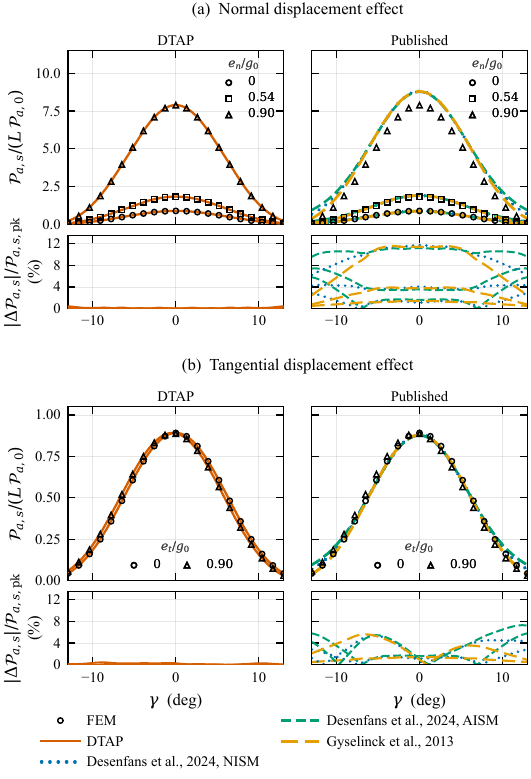}
  \caption{Effect of (a) normal and (b) tangential eccentricity on the skewed air-gap permeance of Motor~2, normalised by the unskewed $\mathcal{P}_{a,0}$.}
  \label{fig:skew}
\end{figure}

Fig.~\ref{fig:skew} repeats Fig.~\ref{fig:eccentricity} for the skewed rotor. The \ac{FEM} reference applies (\ref{eq:P_as_orig}) numerically to a cubic spline through the unskewed \ac{FEM} permeance, converged to 0.02\,\% of the aligned value. The \ac{NISM} and \ac{AISM} are skewed as published~\cite{desenfans2024}, and the method of Gyselinck and Sabariego by the trapezoidal rule over 11 slices. Deviations are again given relative to the aligned \ac{FEM} value of each rotor pose. Skew effectively averages the deviations along $\gamma$: without eccentricity, the mean deviation of \ac{DTAP} decreases from 0.46\,\% to 0.04\,\%, and those of the \ac{NISM} and \ac{AISM} from 2.88--4.27\,\% to 1.33--2.32\,\%, whereas that of Gyselinck and Sabariego increases from 1.01\,\% to 1.05\,\%. At $e_n=0.9\,g_0$, the aligned skewed \ac{FEM} permeance rises by a factor of 8.85, whereas the published methods again scale the entire curve by $h_\mathrm{T}=10$ and deviate by 8.53--10.64\,\%, against 0.12\,\% for \ac{DTAP}. At $e_t=0.9\,g_0$ by 3.06--3.68\,\%, against 0.21\,\%. Hence, the closed-form skew preserves the accuracy of \ac{DTAP} under eccentricity.

\begin{table}[!t]
  \centering
  \caption{Computation time of the skewed air-gap permeance and its partial derivatives.}
  \label{tab:timing}
  \setlength{\tabcolsep}{2.5pt}
  \footnotesize
  \begin{tabular}{@{}lrrrrr@{}}
    \toprule
    Model & $\mathcal{P}_{a,s}$ & $\partial\mathcal{P}_{a,s}/\partial\gamma$ & $\partial\mathcal{P}_{a,s}/\partial e_n$ & $\partial\mathcal{P}_{a,s}/\partial e_t$ & Total \\
    \midrule
    Silva et~al., 2023           & 75.0 & 155 & 3.09 & N/A & 233 \\
    Gyselinck et~al., 2013       & 20.1 & 38.8 & 2.73 & N/A & 61.6 \\
    Desenfans et~al., 2024, NISM & 17.6 & 35.8 & 3.06 & N/A & 56.5 \\
    Ostovi\'{c}, 1989            & 8.01 & 11.4 & 0.73 & N/A & 20.2 \\
    DTAP (this work)             & 15.8 & \textbf{1.04} & 0.96 & 1.93 & 19.8 \\
    Desenfans et~al., 2024, AISM & \textbf{7.26} & 7.42 & \textbf{0.27} & N/A & \textbf{14.9} \\
    \bottomrule
  \end{tabular}

  \smallskip
  \parbox{\linewidth}{\raggedright Derivative times assume $\mathcal{P}_{a,s}$ is given. Unit in $\mu$s. N/A: no dependence on $e_t$. Lowest value per column in bold.}
\end{table}

Table~\ref{tab:timing} lists the corresponding computation times. Silva \textit{et al.} skew by multi-slicing~\cite{silva2023}, and the same 11 slices are applied to the method of Gyselinck and Sabariego, which raises their permeance evaluation to 20.1--75.0\,$\mu$s. The moments raise that of \ac{DTAP} from 6.74 to 15.8\,$\mu$s, within the 7.26--17.6\,$\mu$s of the other closed forms. Its derivative to $\gamma$ (\ref{eq:skew_dgamma}), however, requires no moments, as it only evaluates the integrand at $u\pm\Delta$, and takes 1.04\,$\mu$s, against 7.42--155\,$\mu$s for the published methods. Its derivative to $e_n$ (\ref{eq:skew_den}) reuses the moments and takes 0.96\,$\mu$s, against 2.73--3.09\,$\mu$s for the central differences and 0.27--0.73\,$\mu$s for the closed forms of the \ac{AISM} and Ostovi\'{c}. In total, \ac{DTAP} requires 19.8\,$\mu$s, comparable to Ostovi\'{c} with 20.2\,$\mu$s, more than the \ac{AISM} with 14.9\,$\mu$s, and less than the other methods with 56.5--233\,$\mu$s.

\subsection{MEC Integration}\label{sec:mec}

\begin{figure}[!t]
  \centering
  % TODO: switch to the annotated fig08_mec_network once it exists.
  \includegraphics[width=\columnwidth]{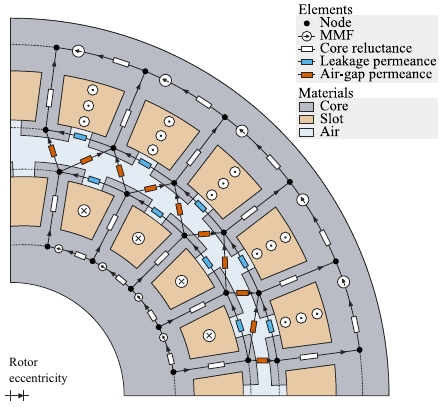}
  \caption{Illustration of used \ac{MEC} of a squirrel-cage \ac{IM}, not to scale. Note that (1) the \ac{MEC} connects every stator tooth to every rotor tooth and (2) it simulates the full circumference of the machine.}
  \label{fig:mec}
\end{figure}

The methods are implemented in the \ac{MEC} of Fig.~\ref{fig:mec}, detailed in~\cite{desenfans2024, gong2025}. Each stator and rotor tooth is represented by a tip node in its head and a yoke node in the yoke behind it, connected by a series of cuboid reluctances, and circumferential hollow-cylinder segment reluctances connect the yoke nodes of neighbouring teeth. These reluctances follow from the nonlinear $B$--$H$ characteristic of pure iron, so that saturation is included. In series with each yoke reluctance, an \ac{MMF} represents the current enclosed by the two neighbouring teeth and their yoke segment, i.e., the turn-weighted phase currents of a stator slot or the bar current of a rotor slot. Slot-opening leakage permeances connect neighbouring tip nodes of the same member, and the air-gap permeances $\mathcal{P}_{a,kj}$ connect every stator tooth $k$ to every rotor tooth $j$. As the core is represented by its own reluctances, $\mathcal{P}_{a,kj}$ need only represent the air gap, in line with the quasi-infinite permeability of the \ac{FEM} identification. The magnetic network is coupled to the electrical equations of the wye-connected stator and of the rotor cage. Skew enters through (\ref{eq:skew_moments}) and the skewed bar length~\cite{desenfans2024}, and the torque and \ac{UMP} follow from (\ref{eq:torque}) and (\ref{eq:ump}).

Motor~2 is supplied at 230\,V line-to-line \ac{RMS} and 60\,Hz, ramped up over the first 500\,ms. It starts unloaded, and its nameplate load of 20.35\,N\,m is applied at 700\,ms. The rotor is held at a static eccentricity of $0.5\,g_0$, and the motor is simulated over 1.233\,s with a time step of 50\,$\mu$s, using the explicit midpoint method and Newton's method for the nonlinear network. The simulation is repeated for each method and, as a reference, with the \ac{FEM} permeance itself, interpolated by cubic splines over $(\gamma,e_n,e_t)$ and integrated along the skew. As all runs share the same supply, load and eccentricity, the air-gap representation is the only difference between them. This reference is preferred over a transient \ac{FEM} model of the complete motor, which would additionally capture effects that the \ac{MEC} neglects or lumps, such as the skin effect in the rotor bars, the leakage flux across the slot bodies and the field distribution within the teeth and yoke segments, so that the air-gap permeance methods could not be compared on their own merit.

\begin{figure}[!t]
  \centering
  \includegraphics[width=\columnwidth]{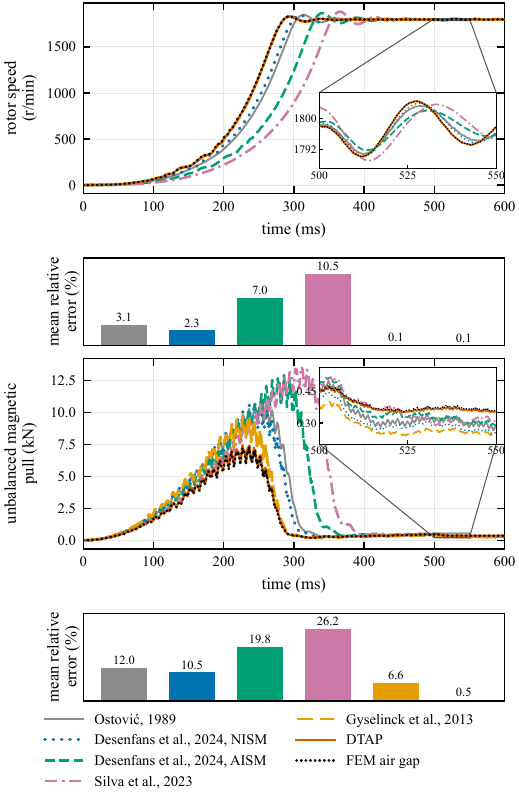}
  \caption{Rotor speed and \acs{UMP} magnitude of Motor~2 during the unloaded run-up at a static eccentricity of $0.5\,g_0$. The bars give the mean absolute deviation from the \acs{FEM} air-gap interpolation, normalised by its peak value.}
  \label{fig:startup}
\end{figure}

During the unloaded run-up in Fig.~\ref{fig:startup}, the reference reaches 95\,\% of its final speed at 277\,ms. \ac{DTAP} reaches it within 0.4\,ms of the reference, with a mean speed deviation of 0.08\,\%, followed by the method of Gyselinck and Sabariego with 0.12\,\%. The other methods delay the run-up by 15--71\,ms, as their peak torque is lower, i.e., 21.9--26.6\,N\,m against 28.3\,N\,m. The \ac{UMP} of the reference peaks at 7.36\,kN at 236\,ms and then decays to a mean of 375\,N between 500 and 600\,ms. \ac{DTAP} reproduces this peak within 1.3\,\% and the settled mean within 1.6\,\%, whereas all published methods overestimate the peak, by 33\,\% for the method of Gyselinck and Sabariego up to 84\,\% for that of Silva \textit{et al.}, and underestimate the settled \ac{UMP} by 12--29\,\%. Even the most accurate published method thus deviates 27 times more than \ac{DTAP} in peak \ac{UMP}. This is expected from (\ref{eq:torque}) and (\ref{eq:ump}): the torque depends on $\partial\mathcal{P}_a/\partial\gamma$, whereas the \ac{UMP} depends on the derivatives to $e_n$ and $e_t$, for which the published methods rely on $h_\mathrm{T}$ and lack a dependence on $e_t$. An accurate torque therefore does not imply an accurate \ac{UMP}.

\begin{figure}[!t]
  \centering
  \includegraphics[width=\columnwidth]{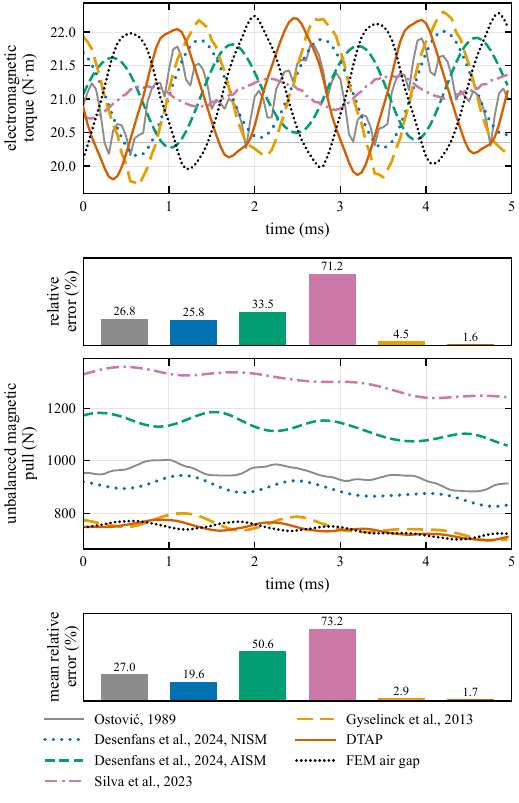}
  \caption{Electromagnetic torque and \acs{UMP} magnitude of Motor~2 at nameplate load and a static eccentricity of $0.5\,g_0$ from 1200\,ms onwards. The upper bars give the deviation of the peak-to-peak torque ripple from that of the \acs{FEM} air-gap representation; the lower bars the mean absolute deviation of the \acs{UMP} relative to the peak of the \ac{FEM} reference.}
  \label{fig:torque_ump}
\end{figure}

At nameplate load, in Fig.~\ref{fig:torque_ump}, all methods settle at 1757\,r/min, so the mean torque is imposed by the load, but its ripple discriminates between the methods. The reference ripple amounts to 2.51\,N\,m peak-to-peak. \ac{DTAP} reproduces it within 1.6\,\%, whereas the method of Gyselinck and Sabariego, the most accurate published method, overestimates it by 4.5\,\%, i.e., a 2.8 times larger deviation, and the other methods smooth it by 25.8--71.2\,\%. As the rotors of \ac{DTAP} and of the method of Gyselinck and Sabariego trail the reference by 4.3$^\circ$ and 7.1$^\circ$ after the run-up, their ripple is shifted in time, so the ripple is compared by its amplitude rather than pointwise. The mean \ac{UMP} rises to 733\,N under load. \ac{DTAP} matches it within 0.2\,\% and the method of Gyselinck and Sabariego within 1.3\,\%, whereas the other methods overestimate it by 19.9--75.2\,\%. As all published methods underestimate the \ac{UMP} at no load, their error even changes sign with the operating point.

\begin{table*}[!t]
  \centering
  \caption{Comparison of steady-state simulation results at rated load and 50\% rotor eccentricity.}
  \label{tab:steady}
  \setlength{\tabcolsep}{8pt}
  \footnotesize
  \sisetup{detect-weight, detect-family}
  \begin{tabular}{@{}l
      S[table-format=1.2]
      S[table-format=4.1]
      S[table-format=2.2]
      S[table-format=2.1]
      S[table-format=1.3]
      S[table-format=4.0]
      S[table-format=1.3]
      S[table-format=3.1]@{}}
    \toprule
    Model & {\shortstack{$T$\\RMS ripple}} & {$\overline{|F|}$} & {$\overline{\angle F}$} & {\shortstack{$|F|$\\RMS ripple}} & {$s$} & {$P_\text{in}$} & {$\mathrm{PF}$} & {$t_\text{step}$} \\
          & {(N$\cdot$m)} & {(N)} & {(deg)} & {(N)} & {(\%)} & {(W)} & {} & {(ms)} \\
    \midrule
    Silva et~al., 2023                & 0.16 & 1283.1 & 2.70 & 43.6 & 2.517 & 4171 & 0.837 & 8.7 \\
    Desenfans et~al., 2024, AISM      & 0.49 & 1111.6 & 3.85 & 39.7 & 2.486 & 4177 & 0.828 & 3.1 \\
    Ostovi\'{c}, 1989                 & 0.41 & 934.8 & 5.37 & 33.5 & 2.452 & 4180 & 0.823 & 3.1 \\
    Desenfans et~al., 2024, NISM      & 0.56 & 878.8 & \bfseries 5.48 & 31.7 & 2.439 & 4177 & 0.830 & \bfseries 2.9 \\
    Gyselinck et~al., 2013            & 0.78 & 742.0 & 7.15 & 28.2 & 2.412 & 4178 & 0.828 & 7.4 \\
    DTAP (this work)                  & \bfseries 0.76 & \bfseries 733.8 & 5.53 & \bfseries 20.8 & \bfseries 2.405 & \bfseries 4174 & \bfseries 0.833 & 3.6 \\
    \midrule
    FEM air gap                       & 0.76 & 732.5 & 5.50 & 20.5 & 2.405 & 4174 & 0.833 & 12.3 \\
    \bottomrule
  \end{tabular}
\end{table*}

Table~\ref{tab:steady} summarises the steady state over twelve electrical periods, with the ripples given as the \ac{RMS} deviation from the mean, $\overline{\angle F}$ the \ac{UMP} direction relative to the eccentricity, $s$ the slip and PF the total power factor. The slip, input power and power factor of all methods lie within 4.7\,\%, 0.2\,\% and 1.2\,\% of the reference, respectively, and \ac{DTAP} matches them to the printed precision, i.e., 2.405\,\% slip, 4174\,W and a power factor of 0.833, whereas the mean \ac{UMP} spans a factor of 1.75 between the methods, from 733.8\,N for \ac{DTAP} to 1283.1\,N for the method of Silva \textit{et al.}, against 732.5\,N for the reference. The air-gap model thus governs the predicted forces far more than the terminal behaviour of the motor. \ac{DTAP} also matches the direction of the \ac{UMP}, with 5.53$^\circ$ against 5.50$^\circ$, as do the \ac{NISM} and the method of Ostovi\'{c}, whereas the other methods deviate by 1.65--2.80$^\circ$. Further, \ac{DTAP} reproduces the torque and \ac{UMP} ripples within 1.5\,\%, whereas the published methods exaggerate the latter by 38--113\,\%. As the closed-form air-gap permeances take only 12--14\,\% of $t_\mathrm{step}$, their cost differences fall within the run-to-run variation; only the slice-based skew of the methods of Silva \textit{et al.} and Gyselinck and Sabariego dominates $t_\mathrm{step}$.

\section{Limitations and Future Work}\label{sec:limitations}

Like all methods that connect teeth by a single permeance, \ac{DTAP} assumes a single magnetic scalar potential per tooth head surface, which its identification enforces through the quasi-infinite permeability of the tooth heads. When the tooth tips saturate, the potential varies across the tooth surface and the air-gap permeance changes, which a saturation-invariant permeance cannot represent. This is a fundamental limitation of this line of research. A related limitation concerns closed slots, as used in many squirrel-cage rotors, whose tooth heads are joined by thin iron bridges that saturate already at low flux. With the quasi-infinite permeability of the identification, these bridges would short-circuit neighbouring teeth, so that \ac{DTAP} applies to closed slots only if the bridges are excluded from its identification and represented by separate nonlinear reluctances in the \ac{MEC}.

An intermediate solution is to identify \ac{DTAP} using \ac{FEM} with saturated tooth heads, suitable for simulating a specific operating point. Alternatively, a future variant of \ac{DTAP} could assign multiple air-gap permeances to each tooth head, connected to separate nodes along its surface, to accommodate the variation of the magnetic scalar potential. Mesh-based \acp{MEC} already include local saturation~\cite{amrhein2009, amrhein2010}, albeit at a significantly higher computational cost.

\section{Conclusion}\label{sec:conclusion}
This work proposed \ac{DTAP}, a closed-form air-gap permeance model for \acp{MEC} of rotor-eccentric electric motors, which sums a flux-tube and a fringing term, each shaped by an inverted cubic Hermite smoothstep. Normal eccentricity scales the two terms differently, tangential eccentricity shifts and tilts the curve, and rotor skew is included in closed form. Its eight parameters are identified from a 2-D \ac{FEM} model of the complete ring of tooth heads, which includes the influence of the neighbouring teeth.

Over three motors, \ac{DTAP} reproduces the \ac{FEM} air-gap permeance with a mean relative error of 0.62--0.78\,\%, which is 2.8--4.8 times lower than that of the most accurate published method, mainly owing to its treatment of eccentricity, at a comparable computational cost. In the \ac{MEC} of an eccentric squirrel-cage \ac{IM}, the choice of air-gap model barely affects the slip, input power and power factor, but can delay the run-up by up to 71\,ms, smooth the torque ripple by up to 71\,\% and shift the mean \ac{UMP} under load by up to 75\,\%. \ac{DTAP} reproduces the run-up within 0.4\,ms, the torque ripple within 1.6\,\% and the mean \ac{UMP} under load within 0.2\,\%, whereas the most accurate published method overestimates the peak \ac{UMP} during the run-up by 33\,\%. \ac{DTAP} thus enables a \ac{MEC} to predict the torque pulsations and bearing loads of eccentric motors at a low computational cost. The effect of tooth-tip saturation and closed slots on the air-gap permeance remains the main direction for future work.

\appendices
\section{Moments of the Smoothstep Function}\label{app:moments}
As $S$ depends on $v$ only through $|v|$, $S(v)$ is even and $v\,S(v)$ is odd, so that $M_0(-x)=-M_0(x)$ and $M_1(-x)=M_1(x)$, and it suffices to consider $x\geq0$. The plateau $0\leq v\leq p$, with $p=\min(x,B)$, contributes $p$ to $M_0$ and $\tfrac{1}{2}p^2$ to $M_1$. On the shoulder, the substitution $v=B+w\varphi$, with $w=A-B$, maps $B<v<A$ to $0<\varphi<1$, where $S=1-3\varphi^2+2\varphi^3$, and the upper limit $v=x$ to $\varphi=f$. The clipping of $f$ makes this contribution vanish for $x\leq B$ and cover the whole shoulder for $x\geq A$. With
\begin{equation}
G_i(f)=\int_0^f\varphi^i\left(1-3\varphi^2+2\varphi^3\right)\mathrm{d}\varphi,\quad i\in\{0,1\},
\end{equation}
the shoulder contributes $w\,G_0(f)$ to $M_0$ and $w\int_0^f(B+w\varphi)(1-3\varphi^2+2\varphi^3)\,\mathrm{d}\varphi=w\big[B\,G_0(f)+w\,G_1(f)\big]$ to $M_1$. Adding both contributions gives (\ref{eq:moments_closed}).

\section*{Acknowledgment}
This work was supported by the Flanders Research Foundation (Fonds voor Wetenschappelijk Onderzoek, FWO), grant number 1298627N.

\bibliographystyle{IEEEtran}
\bibliography{IEEEabrv,references}

\end{document}